\documentclass{IOS-Book-Article}

\usepackage{mathtools, mathptmx, amsmath, amsthm, amssymb}
\usepackage{graphicx, subcaption}
\usepackage{listings}
\usepackage{caption}
\usepackage{multirow}
\usepackage{color}
\usepackage{cite, hyperref}
\usepackage{siunitx}
\usepackage{xspace}
\usepackage{makecell}
\usepackage{textcomp}
\usepackage{xfrac}

\definecolor{gray}{gray}{0.3}
\definecolor{graybox}{gray}{0.8}

\newcommand{\swift}{{\sc Swift}\xspace}
\newcommand{\velociraptor}{{\sc VELOCIraptor}\xspace}

\def\hb{\hbox to 10.7 cm{}}

\begin{document}

\lstset{%
    language=c,
    basicstyle=\scriptsize\ttfamily,
    keywordstyle=\bfseries,
    commentstyle=\color{gray},
    rulecolor=\color{black},
    framerule=0.6pt,
    numbers=left,
    numberstyle=\tiny,
    escapeinside={@}{@},
    captionpos=b
  }

\pagestyle{headings}
\def\thepage{}

\begin{frontmatter}              

\title{A Hybrid MPI+Threads Approach to Particle Group Finding Using Union-Find}

\markboth{}{September 2019\hb}

\author[A]{\fnms{James S.} \snm{Willis}%
\thanks{Corresponding Author; E-mail:\url{james_willis@hotmail.co.uk.}}},
\author[A,B]{\fnms{Matthieu} \snm{Schaller}},
\author[C]{\fnms{Pedro} \snm{Gonnet}}
and
\author[A]{\fnms{John C.} \snm{Helly}},

\runningauthor{Institute for Computational
Cosmology (ICC),
Department of Physics,
Durham University,
Durham DH1 3LE, UK}
\address[A]{Institute for Computational
Cosmology (ICC),
Department of Physics,
Durham University,
Durham DH1 3LE, UK}
\address[B]{Leiden Observatory, 
PO Box 9513, 
NL-2300 RA Leiden,
The Netherlands}
\address[C]{Google AI Switzerland GmbH,
8002 Z{\"u}rich, Switzerland}

\begin{abstract}
The Friends-of-Friends (FoF) algorithm is a standard technique used in
cosmological $N$-body simulations to identify structures. Its goal is to find clusters of particles (called
groups) that are separated by at most a cut-off radius. $N$-body
simulations typically use most of the memory present on a node, leaving
very little free for a FoF algorithm to run on-the-fly. We propose a new
method that utilises the common Union-Find data structure and a hybrid MPI+threads approach. The
algorithm can also be expressed elegantly in a task-based formalism if such
a framework is used in the rest of the application. We have implemented our
algorithm in the open-source cosmological code, \swift. Our implementation displays excellent strong- and
weak-scaling behaviour on realistic problems and compares favourably
(speed-up of 18x) over other methods commonly used
in the $N$-body community.
\end{abstract}

\end{frontmatter}
\markboth{J.~Willis, M.~Schaller, P.~Gonnet~\&~J.~Helly 2019
  \hb}{Hybrid MPI+Thread Approach to Particle Group Finding\hb}

\section{Introduction}
\label{ref:introduction}

Over the last four decades cosmological simulations have been the main tool
used by physicists to confront their theoretical predictions to
observations. By creating more-and-more realistic universes they have 
been able to revolutionise our understanding of
the cosmos and establish the current cosmological model. These simulations
typically involve the evolution of large numbers of particles or resolution
elements under the laws of gravity and hydrodynamics. Given the large
volumes simulated and the ever-growing need for more details, these
simulations are often at the forefront of research in HPC and require
ever-increasing computing capabilities. For instance, the current record
holder, the \emph{Euclid flagship simulation}\cite{ref:PKDGRGAV3}, evolved
$8\times10^{12}$ particles from the Big Bang to the present day and
generated peta-bytes of data.

Putting aside the question of running such simulations, analysing these
large volumes of data poses huge computational challenges as even the most
basic operations require sizeable facilities to simply host the data in
memory. One of the most-widely used post-processing tool for such
simulations is the Friends-of-Friends (FoF) method \cite{ref:FOF_algorithm},
which is designed to identify \emph{groups} of particles that are within a
certain \emph{linking-length}, $l_x$, of each other. If the linking-length is
chosen to be small enough then the method will identify groups that
correspond to structures of particles that have formed due to gravity and 
hence capture information about the evolution of the
Universe. More specifically two particles are in the same \emph{group} if
they are at a distance smaller than $l_x$ of each other. Particles can be
linked to multiple other particles and all particles linked in this way are
in the same group \footnote{More mathematically, the problem can be expressed as determining the connected components of a graph $\mathcal{G}$, based on a set of points $P$, where $\mathcal{G}$ is defined as $\mathcal{G} = (P, E)$ with the set $E = \{\{u,v\}:dist(u,v)\leqslant l_x\} $ and $u,v\in P$.}. The size of a group is later defined as the number of
particles that are linked to each other by this criterion. Particles
without any neighbours within $l_x$ form a group of size one. Since
producing catalogs of particle groups in post-processing can be
prohibitively expensive, in terms of i/o at least, it is common practice to
apply the FoF method on-the-fly at fixed time intervals over the course of the
$N$-body simulation. This also allows the production of FoF outputs at a higher frequency. Over the years many dedicated stand-alone FoF packages have been implemented, recent examples used in
production runs include \cite{ref:VELOCIraptor, ref:FOF_Creasey, ref:scalable_FOF}. Nevertheless, the challenge of efficiently distributing the method over large numbers of nodes on-the-fly, i.e. \emph{whilst reusing the pre-existing data structures put in place for the $N$-body solver}, still remains.

In this paper we present a FoF implementation that exploits the hybrid
shared/distributed parallelism built into the \swift cosmological code
\footnote{See also \url{www.swiftsim.com}.} \cite{ref:SWIFT_PASC, 
ref:SWIFT_SIAM} to achieve excellent efficiency whilst also being able to
run at regular intervals over the course of large cosmological simulations.

\section{FoF using the Union-Find algorithm}
\label{sec:union-find_implementation}

FoF is related to the more general problem of Euclidean minimum spanning trees 
(here in 3 dimensions), which is a very well-studied problem (e.g. \cite{ref:EMST, ref:STANN})
with algorithms that are near-linear in the worst case, but differs crucially in that:
\begin{itemize}
    \item The maximum Euclidean distance considered is limited, thus limiting the range of neighbours for each node, and
    \item We are not interested in the exact structure of the resulting minimum spanning tree (or set of trees), but only in which nodes belong to the same trees.
\end{itemize}

The problem is therefore equivalent to the \em{disjoint-set union} (or \em{union-find}) problem \cite{ref:Galil1991,ref:Union-Find_algorithm}, and the FoF method we have implemented is based on the approaches used for its solution in shared/distributed-memory parallel settings \cite{ref:wait-free_parallel,ref:scalable_union_find,ref:FOF_MPI,ref:Jayanti_2016}.

A disjoint-set data structure is the basis
for the algorithm, which maintains a collection of dynamic non-overlapping sets
consisting of $N$ distinct elements. Each set is identified by a
representative element (the \emph{root}). It is widely used in the
calculation of minimum spanning trees in graphs and the computation of
connected components.

The Union-Find algorithm is designed around two operations: \emph{Union}, 
which merges a pair of sets and \emph{Find}, which identifies the set a given 
element resides in. The data structure is typically implemented using a {\it
  forest}, where each {\it tree} represents a connected set and the root of
each tree identifies the set. Initially each set contains one element which
is the sole member of its set and its set's representative. Two sets
containing elements that are within the linking-length distance, $l_x$, are merged
using the \emph{Union} operation\footnote{In the context of the FoF method
  we use the following terminology: a \emph{set} is referred to as a
  \emph{group} and an \emph{element} is an individual \emph{particle}.}.

There are several standard ways to optimise the
Union-Find algorithm. The \emph{Union} operation for example, can be
implemented using \emph{Union-by-size} which links smaller sets to larger
ones and \emph{Union-by-rank} that links sets with shorter trees to sets 
which have taller trees. However, we will use \emph{Union-by-root} and make the
larger root always point to the smaller root, where the initial root of each set 
is assigned by its offset in the array. This allows us to bypass the issues 
with parallelism (see below) reported by \cite{ref:Jayanti_2016}.

Another common optimisation technique is \emph{path compression}. Each
tree vertex traversed in a Find operation is set to point to the root of the
set. This means that subsequent Find operations are quicker as most vertices
will point directly to the root; reducing the rank of each particle and hence 
lowering the (theoretical) loss of performance using a Union-by-root approach over
a Union-by-rank.

The Union-Find algorithm has been extensively parallelised in the
literature for both shared and distributed memory machines:
\cite{ref:wait-free_parallel, ref:scalable_union_find, ref:FOF_MPI}. The
novelty of our paper is the introduction of a hybrid shared/distributed
memory algorithm that uses a task-based framework, which can be run
on-the-fly within our $N$-body code that imposes a spatial decomposition.

\section{Implementation in the \swift code}
\label{sec:implementation}

\subsection{Serial implementation}
\label{sec:serial}

\begin{figure}
  \centering
  \includegraphics[width=0.6\textwidth]{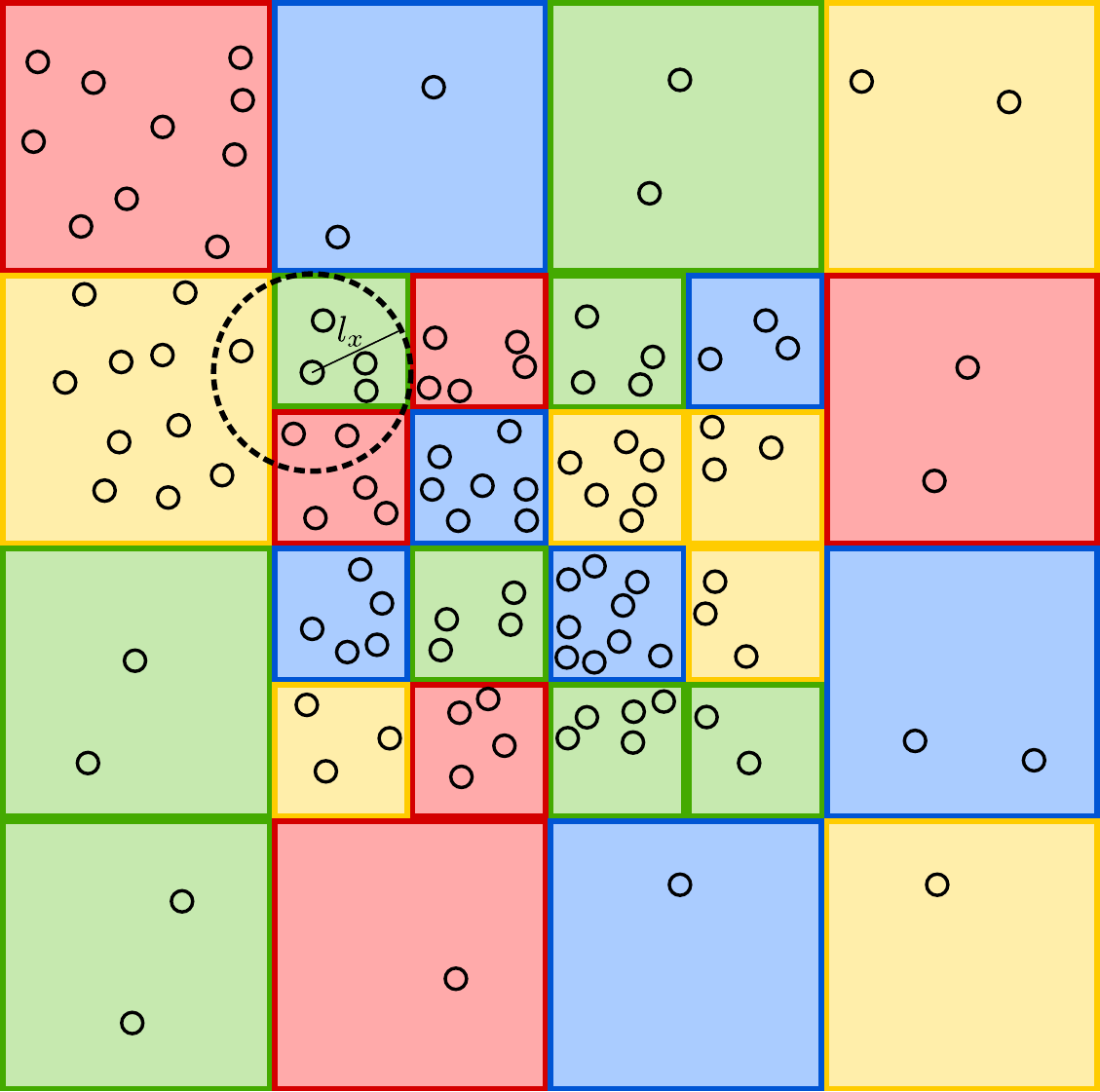}
  \caption{FoF Union-Find using task-based parallelism. Each coloured cell
    represents a single task. ‘Self’ tasks are created for each cell and
    ‘pair’ tasks are created between cells that lie within the cut-off
    radius, $l_x$, of each other. A self task performs a FoF search on
    particles in a single cell whereas a pair task carries out a FoF search
    between particles in neighbouring cells. Tasks are placed into a
    queue. A group of threads pick and execute tasks from the queue
    concurrently until there are none remaining.}
   \label{fig:fof_tasks}
\end{figure}

In practice the Union-Find data structure is implemented using an array of
length, $N$, where $N$ is the total number of particles and each element
represents a particle. The array is initialised so that each particle
exists in its own group, i.e each element is set to the offset of the
particle in the array. A neighbour search is then performed over the
particles using the linking-length, $l_x$, as the search criterion. The
Find operation is used on all particles that are neighbours to return their
roots. Two groups are then merged using the Union operation, where the
smaller of the two roots is used as the group label henceforth. For
example:

\begin{minipage}{\linewidth}
\begin{lstlisting}[caption={Union-Find with a simple iteration over neighbours.}, label=code:naive_pair]
for (int i = 0; i < N; i++)
  for (int j = 0; j < N; j++)
    if(i == j) continue; // Avoid self
    r = particle_dist(parts[i], parts[j]);
    if (r < l_x)
      // Find operation
      int root_i = fof_find(i, group_index);
      int root_j = fof_find(j, group_index);
      // Union operation
      if(root_i < root_j) group_index[root_j] = root_i;
      else group_index[root_i] = root_j;
\end{lstlisting}
\end{minipage}

\noindent where {\tt parts} is the particle array and {\tt group\_index} is
the array that represents the Union-Find data structure.

As in the case of minimum spanning tree problems, we make use of the octree 
(quadtree in 2D) present in \swift to
significantly reduce the cost of the neighbour search, by
only recursing on pairs of cells that are within the requested cut-off
radius, $l_x$, of each other. We note, however, that the best performance is
achieved when the size of the tree nodes matches the linking-length (see
the technique of \cite{ref:FOF_Creasey} or \cite{ref:scalable_FOF}), but that tailoring the
octree node sizes would hinder the performance of the rest of the \swift
code and is hence not an option. Once the tree has been setup, the problem
becomes almost embarrassingly parallel and we split the workload
evenly either between: (a) a group of threads, or equivalently (b) a set
of tasks (see Fig. \ref{fig:fof_tasks}). We implement the latter in \swift
using a variant of the {\sc QuickSched} tasking library
\cite{ref:QuickSched}.

\subsection{Shared memory parallelism}
\label{sec:shared_memory}

In order to parallelise the algorithm a subtle issue needs to be taken care
of, i.e. each thread must have a consistent view of the tree data. For example,
consider two roots: $r_i$ and $r_j$, we need to ensure that one thread does
not find $r_i < r_j$ whilst another concludes $r_i > r_j$. One
possibility would be to use locks when writing to the Union-Find data structure ({\tt group\_index}), but this
would hinder scalability as more and more threads try to access the list.
We instead solve this problem by checking that the value of $r_i$ has not
changed between being read and being found to be lower than $r_j$. If $r_i$
has changed between these events the process is repeated until the value of
$r_i$ remains constant.

We implement the Union operation in a thread-safe manner by using the
\emph{Compare And Swap} (CAS) atomic, proposed by \cite{ref:wait-free_parallel}:

\begin{minipage}{\linewidth}
\begin{lstlisting}[caption={Using a CAS atomic operation to perform the \emph{Union} of two groups in a thread-safe manner.}, label=code:cas_union]
int atomic_update_root(volatile size_t *address, size_t y) {

  size_t *size_t_ptr = (size_t *)address;

  size_t old_val = *address;
  size_t test_val = old_val;
  size_t new_val = y;

  old_val = atomic_cas(size_t_ptr, test_val, new_val);

  return (test_val == old_val);
}

void fof_union(size_t i, size_t j, size_t *group_index) {
  int result = 0;
  // Loop until the root can be set to a new value.
  do {
    size_t root_i = fof_find(i, group_index);
    size_t root_j = fof_find(j, group_index);

    if(root_j < root_i)
      result = atomic_update_root(&group_index[root_i], root_j);
    else 
      result = atomic_update_root(&group_index[root_j], root_i);
  
  } while (!result);
}
\end{lstlisting}
\end{minipage}

\noindent This ensures any update to {\tt group\_index} is
lock-free, and hence avoids any performance penalties introduced by locks. A weakness of this method, however, is that the CAS operation can only update a single variable at a time\footnote{There has been an attempt by \cite{ref:double_cas} to implement a multi-variable CAS operation, but their results show that in practice the performance of this approach is not superior to traditional locking techniques.}. Therefore, if a (formally more efficient) Union-by-size or Union-by-rank version of the algorithm were to be used, it would require a lock instead of an atomic to avoid data races. One solution to this problem is to adopt the approach by \cite{ref:Jayanti_2016}, where the Union is instead randomised. It avoids having to update two variables per Union as the size or rank of a group is not stored in addition to the root.

We also tested a version of our parallel algorithm using the randomisation technique proposed by \cite{ref:Jayanti_2016}. This implementation showed similar times to solution compared to our basic approach. This is due to the fact that we only use the root of a group to perform a Union operation, and hence do not suffer from the weakness of the Anderson \& Woll implementation  \cite{ref:wait-free_parallel}. Our specific workloads, where the \emph{rank} of the elements added in the \emph{Union} operations are typically small, are another reason why we did not see a noticeable increase in performance. For these reasons we chose to use our simpler solution and stick to the \emph{Union-by-root} method.

\subsection{Distributed memory parallelism}
\label{sec:distributed_memory}

\begin{figure}
  \vspace{-0.2cm}
  \centering \includegraphics[width=0.8\textwidth]{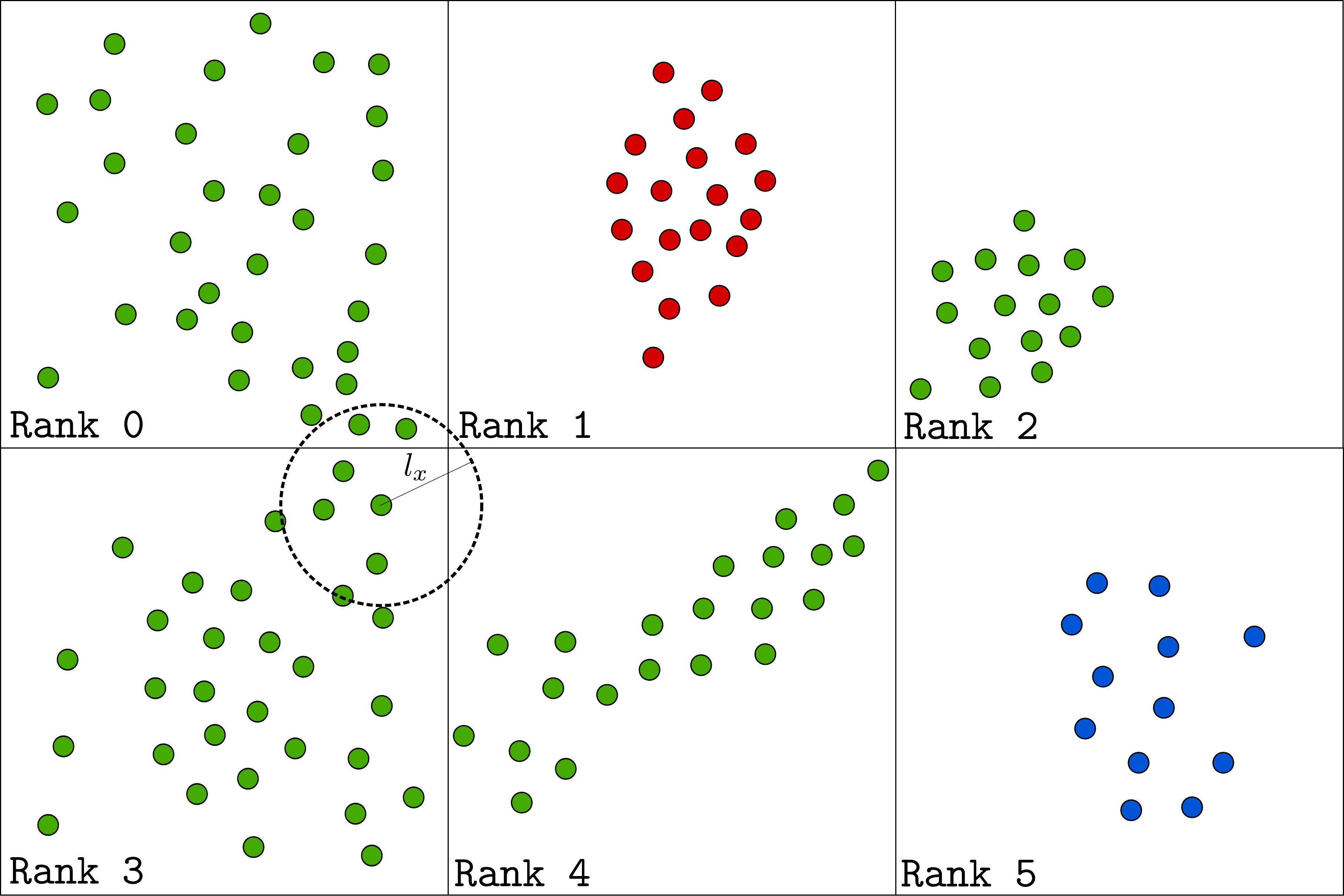}
  \vspace{-0.2cm}
  \caption{Distributed Union-Find over MPI. Particles are distributed
    across each MPI rank and the following steps are performed: 1) a local
    FoF is performed on each MPI rank; 2) relabel group IDs so that they
    are globally unique; 3) identify links between groups that span two MPI
    domains; 4) merge distributed groups and agree on ownership. The figure
    also illustrates an edge case that can occur. The group on rank 0 is
    indirectly linked to the group on rank 2 via the groups on ranks 3 and
    4. If we were to only merge groups between MPI ranks that are direct
    neighbours in step 4), we would fail to take into account this subtlety
    and miss the indirect group links.}
   \label{fig:fof_mpi}
\end{figure}

For larger simulations, particles are distributed across multiple nodes (see Fig. \ref{fig:fof_mpi}). To
address the problem of groups spanning multiple nodes, we follow the
strategy outlined by \cite{ref:FOF_MPI} and improve upon it to handle the
case where the number of groups is much larger than $10^2$.

We first perform a multi-threaded local Union-Find on each node, as
described in Section~\ref{sec:shared_memory}, followed by assigning unique
group IDs across all nodes. This is done by computing an offset based upon
the MPI rank of the node. Each rank, $p$, computes a sum of
the total number of particles contained on every MPI rank lower than
itself, $\scriptstyle\sum\limits_{i=0}^{i<p}N_i$, where $N_i$ is the total number of particles present on rank $i$. The sum is then used to offset all group IDs on the local node. In practice this is done using \texttt{MPI\_Scan}: 

\begin{minipage}{\linewidth}
\begin{lstlisting}[caption={Computing the node offset with \texttt{MPI\_Scan}.}, label=code:node_offset_calc]
  long long num_parts_cumulative;
  long long num_parts_local = num_parts;
  MPI_Scan(&num_parts_local, &num_parts_cumulative, 1, MPI_LONG_LONG, 
           MPI_SUM, MPI_COMM_WORLD);
  size_t node_offset = num_parts_cumulative - num_parts_local;
\end{lstlisting}
\end{minipage}

Next, we identify links between groups that span at least two node domains,
only communicating information for groups that are within the
linking-length, $l_x$, of the domain boundaries. This greatly reduces the
amount of data replication. The final step performs a global gather
communication (\texttt{MPI\_Allgatherv}) on the list of group links so that
every node has access to the global list of group links ({\tt global\_group\_links}). Each node then
applies the Union-Find to {\tt global\_group\_links}, only updating the
roots of groups which are local to them. This ensures that all
spanning groups are merged and each node agrees upon group ownership.

\begin{figure}
  \centering
  \includegraphics[width=0.8\textwidth]{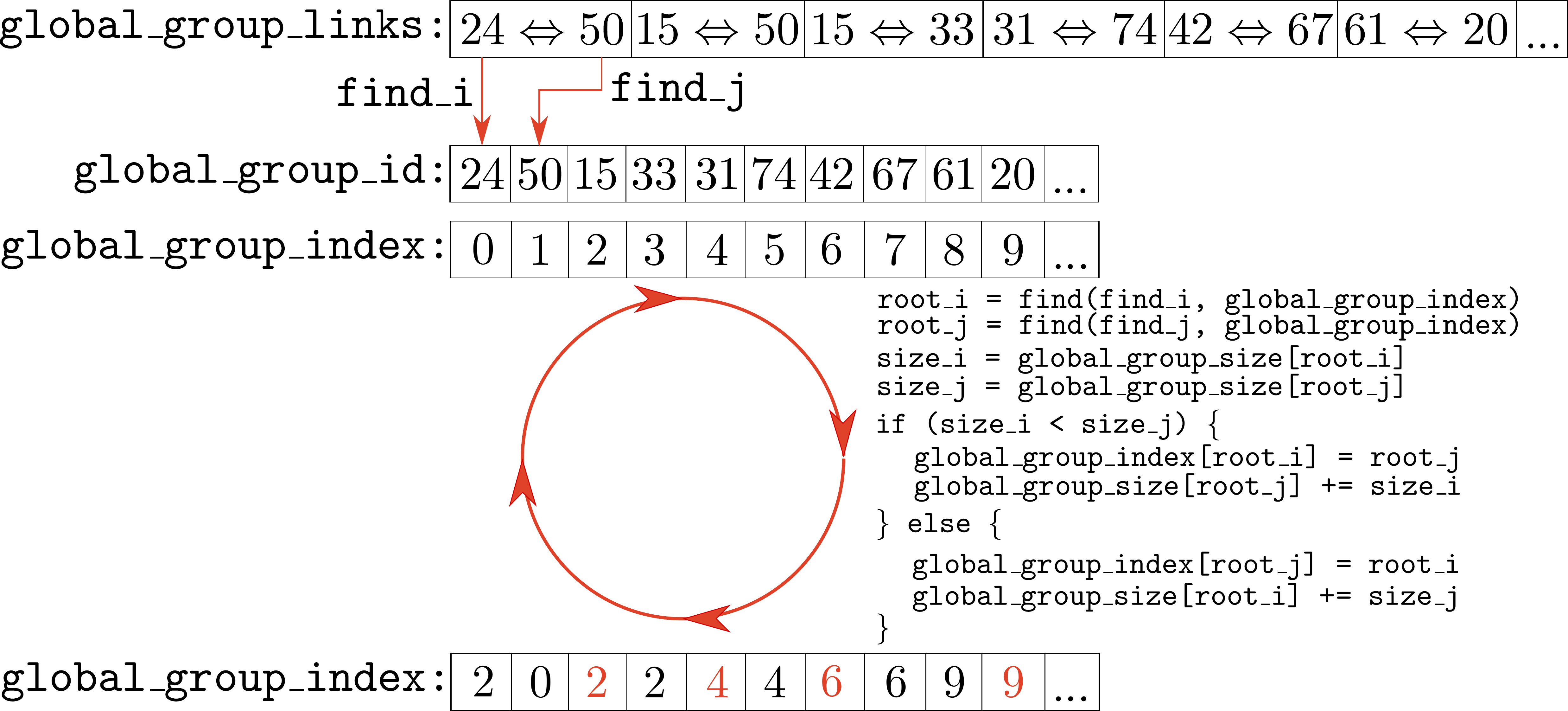}
  \caption{ Step 4 in the distributed
    Union-Find method. Distributed groups are merged and each MPI rank
    agrees on group ownership. The {\tt global\_group\_links} array stores all group
    links that span an MPI domain, each unique ID in the list is unpacked into
    {\tt global\_group\_id} and mapped to a number between 0 and the total
    number of unique group IDs in the list ({\tt
      global\_group\_index}). Find is applied to each pair of links in the
    list, where the group offset ({\tt find\_i} \& {\tt find\_j}) into {\tt
      global\_group\_id} is used as input. This ensures that Find returns
    the correct group ID in the case where it has been updated in an earlier group merger (Union). 
    The pair of groups are then merged using
    the Union operation and {\tt global\_group\_index} is updated.}
   \label{fig:fof_mpi_method}
\end{figure}

In order to apply the Union-Find on the global list we map each group ID to
a number between 0 and the total number of group IDs that span node
domains. The same group ID may appear multiple times in the list, therefore
we need to search for the first occurrence of it and use the index as input
to the Find operation. This ensures that the result of the Find operation
is correct, as the group ID could have previously been updated from a group
merger earlier in the list. See Fig. \ref{fig:fof_mpi_method}.

Naively one may think that each rank need only run the Union-Find on the
group links that it shares with its neighbouring ranks. However,
Fig. \ref{fig:fof_mpi} shows a particle distribution that forms a group on
rank 0 that is indirectly linked to the group on rank 2 via the groups on
ranks 3 and 4. This group linkage will be overlooked if each rank
only searches for links with its direct neighbours.

If we use the same Union strategy as the local FoF, the distribution of
roots of spanning groups will be skewed towards the lower MPI ranks. This
can lead to a load imbalance between nodes when assigning new local roots
during Step 4. To address this problem we use Union-by-size when merging
groups across MPI domains. This creates an even work load between ranks as
Union-by-size will assign roots more arbitrarily and will only be based
upon the domain decomposition.

\subsection{Implementation details}

\subsubsection{Hash table}

Performing the last step of the distributed FoF algorithm can become quite
expensive, as the length of {\tt global\_group\_links} scales with the node
count. This is because searching for the index of a group ID into the list
roughly takes $\mathcal{O}(N^2)$ operations. A hash table on the other
hand has constant look-up times, $\mathcal{O}(1)$. Therefore we
construct a hash table of group IDs in the list and store their index into {\tt global\_group\_links}.

We also make use of a hash table when calculating the group sizes in the
local FoF. To find the group sizes in serial we loop through the {\tt
  group\_index} array and increment {\tt group\_size} indexed by the
root of the group that each particle is in:

\begin{minipage}{\linewidth}
\begin{lstlisting}[caption={Group size calculation in serial.}, label=code:group_size]
  for (int i = 0; i < N; i++)
    group_size[fof_find(i, group_index)]++;    
\end{lstlisting}
\end{minipage}

\noindent In parallel we divide $N$ by the number of threads and have each
thread work on a section of {\tt group\_index}. We avoid race conditions between threads by protecting access to {\tt group\_size}. To do this we use a hash table to store the
group sizes and root of each group. Once we have looped over {\tt
  group\_index}, we pull out each element of the hash table and write the
intermediate group size to the global {\tt group\_size} array using an atomic
addition (for instance GNU C's {\tt \_\_sync\_fetch\_and\_add}).

\subsubsection{Early elimination of small groups}

The majority of groups in cosmological simulations are of lone
particles. We were able to take advantage of this fact to lower the memory footprint significantly
when calculating group sizes. When constructing the hash table only groups of size
$\geqslant 2$ were stored. We achieved this by initialising each
element of the {\tt group\_size} array to 1, which allowed us to exclude
root particles in the hash table as their contribution to the group size
was already accounted for.

\subsubsection{Path compression optimisation}

The Find operation is a tree traversal that retrieves the root of a group for a given particle. Hence, the execution time is dominated by the depth of the tree at each particle. To amortise the cost of this operation we have implemented path compression. But instead of
compressing trees of all depths, we found it was quicker to only compress
trees with a depth of at least 2.

\section{Results}
\label{sec:results}

To test the performance of our FoF implementation we ran a number of
different benchmarks. We measured the strong- and weak-scaling performance
as well as the speed-up over another FoF application. All results were
obtained on the {\sc cosma-7} DiRAC 2.5x ``Memory Intensive'' System,
located at the University of Durham \footnote{The system consists of 452
  nodes of 2 Intel Xeon Gold 5120 CPUs running at 2.2GHz (14 physical cores
  with AVX512 capability) with 512 GBytes of RAM. The nodes are connected
  using Mellanox EDR Infiniband in a 2:1 blocking configuration. The strong
  scaling results were obtained by running on the MAD02 machine at Durham
  with Turbo Boost disabled for the purposes of obtaining accurate
  measurements. It is a quad socket system each with an Intel Xeon Platinum
  8180 CPU running at 2.5GHz (28 physical cores with AVX512 capability)
  with 1.5 TBytes of RAM. See
  \url{https://dirac.ac.uk/resources/\#MemoryIntensive} for more details on
  each system.}. The results are based on version 0.8.2 of \swift
({\tt git} revision {\tt f05bd301}), which implements the algorithm
described in Section~\ref{sec:implementation}.

\subsection{Measurement methodology}
To get a realistic workload, all benchmarks were carried out using particle
data from the flagship EAGLE simulations \cite{ref:eagle} at late times
(redshift $z=0.1$). The input data contains $4.25\times10^8$ particles split into $\sim2\times10^5$ groups of length $>20$. The workload is
representative of an actual production run of \swift and nicely
fits within a single node's memory. To create a
weak-scaling test, we replicate the simulation volume periodically $N$
times along each axis, creating problem sizes that are $N^3$ larger than
the original volume.

We used the Intel compiler and MPI library v.18.0.2\footnote{with the flags
  \texttt{-O3 -xCORE-AVX512}. } as well as the GNU compiler
v.9.1.0\footnote{with the flags {\tt-O3 -ffast-math -march=skylake-avx512
    -mavx512dq}.}. To obtain precise execution times we used the {\tt
  RDTSC} cycle counter and converted the cycle counts to seconds using
the clock-speed of the CPU. Each data point is the average time of 3
independent runs and the standard deviation is used to measure the uncertainty.
For the weak-scaling tests, we use 4 MPI ranks per node (2 per NUMA
region) and use the MPI version of the code even for the single-node data
point in order to have the same MPI-related overheads throughout the
test. The strong-scaling test does not use MPI and hence probes the
efficiency of the shared memory algorithm.

\subsection{Strong- and weak-scaling results}

\begin{figure}
  \centering
  \vspace{-0.5cm}
  \includegraphics[width=1.0\textwidth]{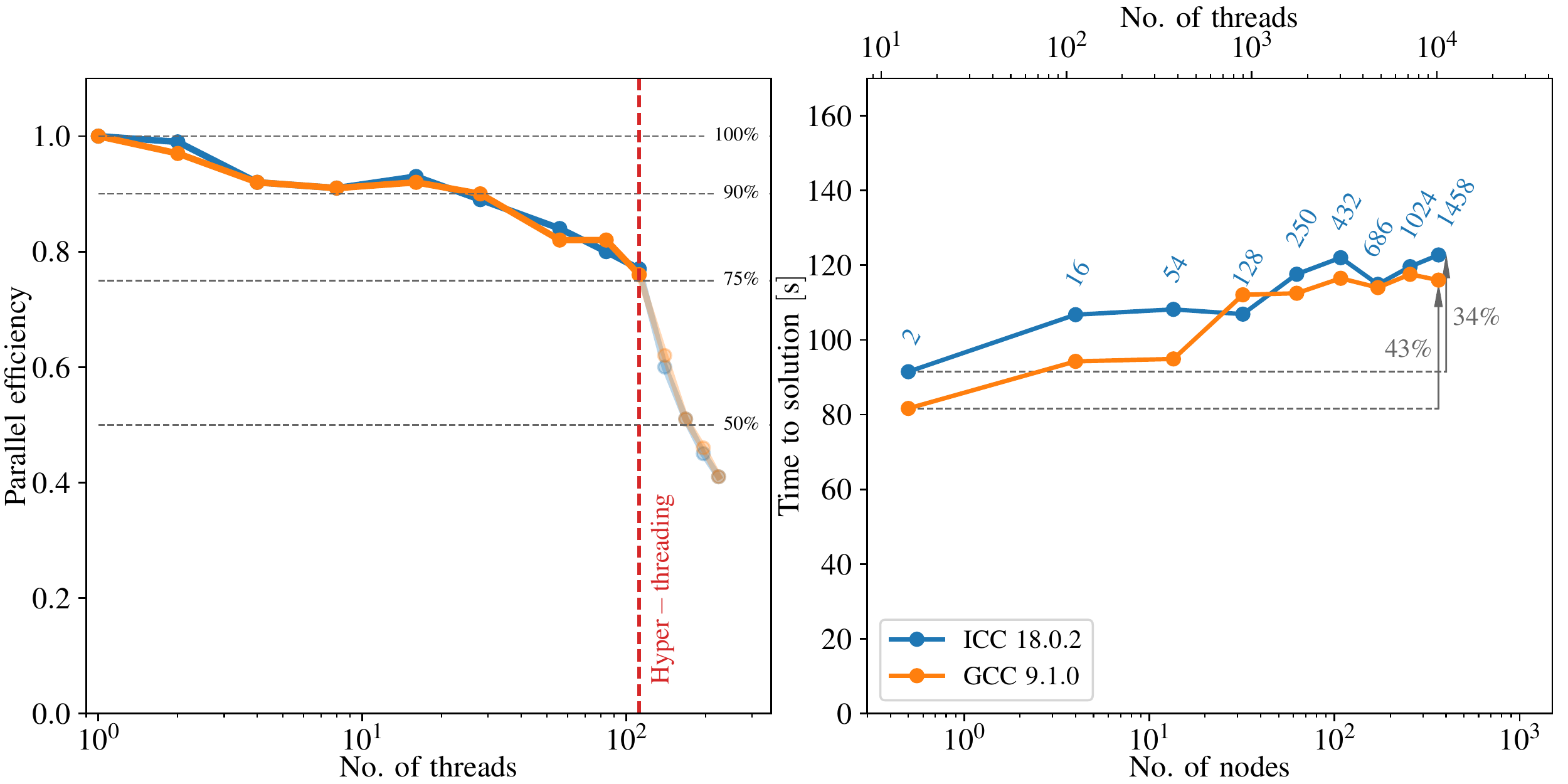}
  \vspace{-0.5cm}
  \caption{\swift FoF scaling results on a representative cosmological
    problem. The particle data is taken
    from the EAGLE simulations \cite{ref:eagle} from a snapshot at redshift
    $z=0.1$, i.e. near the end of the calculation when the distribution of
    particles is far from uniform. ({\it Left}) Strong scaling results. 
    The particle load was kept constant at $\num{4.25e8}$ whilst the number of cores was 
    increased. As the benchmark was performed on one node, the non-MPI
    version of the algorithm was used. We maintain very good strong scaling
    performance and obtain 77\% parallel efficiency on 112 cores. The
    efficiency drops when running with hyper threads due to resource
    contention between threads. ({\it Right}) Weak-scaling results. The
    number of particles per core is kept constant at $\num{3e7}$, as we
    increase the core count. We use 4 MPI ranks per node (2 ranks {\it per
      socket}). For convenience, the total number of MPI ranks used is
    indicated by the labels above the data points. The vertical arrow
    displays the percentage loss in performance running on 10,206 cores,
    which was 43\% for GCC and 34\% for ICC. We achieve good weak scaling from \sfrac{1}{2} a node to
    364\sfrac{1}{2} nodes (a factor of 729 increase in the number of particles and number of cores) despite the overhead costs of MPI communication. For both panels, the standard deviation of each measurement is smaller than the symbol size.}
   \label{fig:strong_weak_scaling}
\end{figure}

The strong scaling results are shown in the left hand panel of
Fig. \ref{fig:strong_weak_scaling}. We stress that 
these results were obtained starting from one core and keeping the problem size
constant. Turbo Boost was also disabled on the node for the purposes of obtaining accurate measurements. We display very good strong scaling and
maintain a high parallel efficiency, achieving 77\% on 112 cores. 
 Only dropping in efficiency when hyper-threads are used, but this can be
explained by resource contention between competing threads. This is a
result of our shared memory strategy: effective load balancing between
threads using an octree and task-based parallelism; and a lock-less
implementation of the parallel Union-Find algorithm.

The right-hand panel displays the weak-scaling performance, where we
achieve good scaling up to 10,206 cores despite the overhead costs of
MPI communication. The last data point corresponds to a simulation with
$3\times10^{11}$ particles. The jump from \sfrac{1}{2} a node to 4 nodes is a result of the
MPI communication being performed over the network, as opposed to on a single node. Additionally, since that data point only
uses half the available cores on the node, a better memory throughput is
achieved and the cores are running at a slightly higher clock speed (2.9 vs. 2.6
GHz) thanks to Turbo Boost. We hence only consider the results starting from
the next data point (4 nodes) where all the cores are busy on each node. From that
point onwards, the gradual increase in runtime is a result of the network,
as it has a greater effect at higher node counts and becomes the limiting
factor. The loss in performance running on 10,206 with ICC is 34\%. Starting
from the second data point (where the nodes are now using all cores and
do not suffer from the caveats mentioned above), we obtain a significant
improvement only losing 15\% going from 4 ($=2^3$ the original problem size) 
to 364\sfrac{1}{2} nodes ($=9^3$ the original problem size). There is also a
noticeable difference in runtime between the Intel and GNU compilers for the first three data points, with GNU showing a speed-up of $\sim$13\% over Intel. A
similar discrepancy is also seen in the strong scaling results.

This is a combination of a highly efficient parallel Union-Find algorithm
within a single node and a scalable distributed memory strategy between
nodes. The domain decomposition implemented in \swift also plays a role so
as to keep the work load balanced between MPI ranks (see
\cite{ref:SWIFT_PASC} and \cite{ref:SWIFT_josh}).

\subsection{Comparison to other software}

As another performance test we compared our implementation against \velociraptor
\cite{ref:VELOCIraptor}, a
FoF application commonly used in the literature. We used the same setup as in the strong-scaling test and ran
on the MAD02 machine using the Intel compiler and MPI library
v.18.0.2\footnote{with the flags \texttt{-O3 -xCORE-AVX512}}. We ran the
non-MPI version of our code and the MPI version of \velociraptor with 1
rank per core. Our FoF took 13.2s to run to completion and \velociraptor took 242s, leading to a net
speed-up of 18.3x\footnote{Note that we used the MPI version of \velociraptor as it was
significantly faster than its shared-memory (OpenMP) version which took 1882s running with 112 threads on the same setup.}. Both codes yield the same answer. Given the large difference in run time on one node and the good weak-scaling displayed by our implementation, we decided not to compare our performance with \velociraptor at scale.

\section{Conclusions}

We presented an efficient and scalable new implementation of the FoF
method that is commonly used to identify structure in cosmological
simulations. The Union-Find data structure was used to create a {\it
 forest} of particles, where each {\it tree} contains a set of particles
that share the same group. A hybrid approach was adopted using threads and
MPI, which allows it to optimally utilise both shared and distributed
memory machines. We made use of atomics to update the Union-Find data structure
 which ensures our implementation remains lock-free. The neighbour search
over particles was sped up using the octree present in the \swift code. A
hash table was used in both the group size calculation and group merging across MPI domains to lower the memory
footprint and improve the time to solution.

When implemented in the \swift code our FoF algorithm achieves good
weak-scaling from 14 to 10,206 cores and displays 
good strong-scaling
performance, maintaining 77\% parallel efficiency running on 112 cores. We
also compare favourably with the commonly used FoF application
\velociraptor, obtaining a speed-up of 18x over it. Together with the weak-scaling performance displayed up to $10^4$ cores this speed-up should allow for an efficient run time when used on-the-fly in production simulations using $\gtrsim 10^5$ cores.

\vspace{-0.5cm}
\section*{Acknowledgements}
\vspace{-0.2cm}
We thank the anonymous referees for their comments that greatly helped improve
the paper. This work would not have been possible without Lydia Heck, Peter Draper, 
Richard Regan and Alastair Basden's help and expertise running on the cosma systems;
as well as the \swift team for their help and input on this project. This work is
supported by {\sc Intel} through establishment of the ICC as an {\sc Intel}
parallel computing centre (IPCC). Matthieu Schaller is additionally supported by the NWO
VENI grant 639.041.749. This work used the DiRAC@Durham facility managed by
the Institute for Computational Cosmology on behalf of the STFC DiRAC HPC
Facility (\url{www.dirac.ac.uk}). The equipment was funded by BEIS capital
funding via STFC capital grants ST/K00042X/1, ST/P002293/1, ST/R002371/1
and ST/S002502/1, Durham University and STFC operations grant
ST/R000832/1. DiRAC is part of the National e-Infrastructure.
\vspace{-0.5cm}



\begin{thebibliography}{10}

\bibitem{ref:PKDGRGAV3}
D.~Potter, J.~Stadel, and R.~Teyssier, ``Pkdgrav3: beyond trillion particle
  cosmological simulations for the next era of galaxy surveys,'' {\em
  Computational Astrophysics and Cosmology}, vol.~4, p.~2, May 2017.

\bibitem{ref:FOF_algorithm}
M.~Davis, G.~Efstathiou, C.~S.~Frenk, and S.~White, ``The evolution of
  large-scale structure in a universe dominated by cold dark matter,'' {\em The
  Astrophysical Journal}, vol.~292, 06 1985.

\bibitem{ref:VELOCIraptor}
P.~J. Elahi {\em et~al.}, ``Hunting for galaxies and halos in simulations with
  velociraptor,'' {\em Publications of the Astronomical Society of Australia},
  vol.~36, p.~e021, 2019.

\bibitem{ref:FOF_Creasey}
P.~E. Creasey, ``{Tree-less 3d Friends-of-Friends using Spatial Hashing},''
  {\em Astron. Comput.}, vol.~25, pp.~159--167, 2018.

\bibitem{ref:scalable_FOF}
Y.~Kwon {\em et~al.}, ``Scalable clustering algorithm for n-body simulations in
  a shared-nothing cluster,'' in {\em Scientific and Statistical Database
  Management}, (Berlin, Heidelberg), pp.~132--150, Springer Berlin Heidelberg,
  2010.

\bibitem{ref:SWIFT_PASC}
M.~Schaller {\em et~al.}, ``Swift: Using task-based parallelism, fully
  asynchronous communication, and graph partition-based domain decomposition
  for strong scaling on more than 100,000 cores,'' in {\em Proceedings of the
  PASC Conference}, PASC 16, (New York, USA), pp.~2:1--2:10, ACM, 2016.

\bibitem{ref:SWIFT_SIAM}
P.~Gonnet, ``Efficient and scalable algorithms for smoothed particle
  hydrodynamics on hybrid shared/distributed-memory architectures,'' {\em SIAM
  Journal on Scientific Computing}, vol.~37, no.~1, pp.~C95--C121, 2015.

\bibitem{ref:EMST}
S.~Arya and D.~M. Mount, ``A fast and simple algorithm for computing
  approximate euclidean minimum spanning trees,'' in {\em Proceedings of the
  Twenty-Seventh Annual {ACM-SIAM} Symposium on Discrete Algorithms, {SODA}
  2016, Arlington, VA, USA, January 10-12, 2016}, pp.~1220--1233, 2016.

\bibitem{ref:STANN}
M.~{Connor} and P.~{Kumar}, ``Fast construction of k-nearest neighbor graphs
  for point clouds,'' {\em IEEE Transactions on Visualization and Computer
  Graphics}, vol.~16, pp.~599--608, July 2010.

\bibitem{ref:Galil1991}
Z.~Galil and G.~F. Italiano, ``Data structures and algorithms for disjoint set
  union problems,'' {\em ACM Computing Surveys (CSUR)}, vol.~23, no.~3,
  pp.~319--344, 1991.

\bibitem{ref:Union-Find_algorithm}
R.~E. Tarjan, ``Efficiency of a good but not linear set union algorithm,'' {\em
  J. ACM}, vol.~22, pp.~215--225, Apr. 1975.

\bibitem{ref:wait-free_parallel}
R.~J. Anderson and H.~Woll, ``Wait-free parallel algorithms for the union-find
  problem,'' in {\em In Proc. 23rd ACM Symposium on Theory of Computing},
  pp.~370--380, 1994.

\bibitem{ref:scalable_union_find}
F.~Manne and M.~M.~A. Patwary, ``A scalable parallel union-find algorithm for
  distributed memory computers,'' in {\em Parallel Processing and Applied
  Mathematics} (R.~Wyrzykowski, J.~Dongarra, K.~Karczewski, and J.~Wasniewski,
  eds.), (Berlin, Heidelberg), pp.~186--195, Springer Berlin Heidelberg, 2010.

\bibitem{ref:FOF_MPI}
C.~Harrison, H.~Childs, and K.~P. Gaither, ``Data-parallel mesh connected
  components labeling and analysis,'' in {\em Proceedings of the 11th
  Eurographics Conference on Parallel Graphics and Visualization}, EGPGV '11,
  (Aire-la-Ville, Switzerland, Switzerland), pp.~131--140, Eurographics
  Association, 2011.

\bibitem{ref:Jayanti_2016}
S.~V. Jayanti and R.~E. Tarjan, ``A randomized concurrent algorithm for
  disjoint set union,'' in {\em Proceedings of the 2016 ACM Symposium on
  Principles of Distributed Computing}, PODC '16, (New York, NY, USA),
  pp.~75--82, ACM, 2016.

\bibitem{ref:QuickSched}
P.~{Gonnet}, A.~B.~G. {Chalk}, and M.~{Schaller}, ``{QuickSched: Task-based
  parallelism with dependencies and conflicts},'' {\em arXiv e-prints},
  p.~arXiv:1601.05384, Jan 2016.

\bibitem{ref:double_cas}
T.~L. Harris, K.~Fraser, and I.~A. Pratt, ``A practical multi-word
  compare-and-swap operation,'' in {\em Distributed Computing} (D.~Malkhi,
  ed.), (Berlin, Heidelberg), pp.~265--279, Springer Berlin Heidelberg, 2002.

\bibitem{ref:eagle}
J.~Schaye {\em et~al.}, ``{The EAGLE project: simulating the evolution and
  assembly of galaxies and their environments},'' {\em Monthly Notices of the
  Royal Astronomical Society}, vol.~446, pp.~521--554, 11 2014.

\bibitem{ref:SWIFT_josh}
J.~{Borrow} {\em et~al.}, ``{SWIFT: Maintaining weak-scalability with a dynamic
  range of $10^4$ in time-step size to harness extreme adaptivity},'' in {\em
  Proceedings of the 13th SPHERIC International Workshop}, SPHERIC 13, pp.~44
  -- 51, Jun 2018.

\end{thebibliography}
\end{document}